\documentclass[twocolumn, prx, aps, superscriptaddress]{revtex4-2}
\usepackage[english]{babel}
\usepackage[utf8]{inputenc}
\usepackage{lmodern}

\usepackage{booktabs} 
\usepackage{amsmath}  
\usepackage{microtype}
\usepackage{braket}
\usepackage{physics}
\usepackage{graphicx}
\usepackage{graphics}
\graphicspath{{figures/}} 
\usepackage{bm}
\usepackage{xcolor}
\usepackage{nicefrac}
\usepackage{hyperref}

\newcommand{\dtilde}[1]{\tilde{\tilde{{#1}}}}

\begin{document} 

\title{Quantum Fisher information in many-photon states from shift current shot noise}

\author{Evgenii Barts}
\email{evgenii.barts@riken.jp}
\affiliation{RIKEN Center for Emergent Matter Science (CEMS), Wako, Saitama 351-0198, Japan}

\author{Takahiro Morimoto}
\affiliation{Department of Applied Physics, The University of Tokyo, Hongo, Tokyo 113-8656, Japan}

\author{Naoto Nagaosa}
\email{nagaosa@riken.jp}
\affiliation{RIKEN Center for Emergent Matter Science (CEMS), Wako, Saitama 351-0198, Japan}
\affiliation{Fundamental Quantum Science Program (FQSP), TRIP Headquarters, RIKEN, Wako 351-0198, Japan}

\date{\today}

\begin{abstract}
Quantum Fisher information~(QFI) sets the ultimate precision of optical phase measurements and reveals multiphoton entanglement, but it is not accessible with conventional photodetection. We theoretically predict that a photodetector utilizing the shot noise of the quantum-geometric shift current of exciton polaritons can directly measure the QFI of nonclassical light. By solving the Lindblad equation, we obtain the time-dependent nonlinear photocurrent for an arbitrary initial photon state. It turns out that, regardless of the quantum state of the incident light, the integrated current depends only on the mean photon number. 
In stark contrast, the shot noise retains the quantum information: its Fano factor is proportional to the photon number variance and therefore encodes the QFI. Numerical calculations confirm these relations for illumination with optical Schr\"{o}dinger cat and squeezed vacuum states. Quantum correlations in nonclassical light, usually hidden from direct detection, become observable in the form of shift current shot noise.

\end{abstract}
\maketitle


\noindent
Quantum photonics, i.e., the science and engineering of quantum states of light for 
quantum communication, quantum computation, and quantum sensing,
is one of the hot topics attracting much attention 
\cite{10.1063/5.0201107,PRXQuantum.5.010101}. 
One fascinating example is the discovery of gravitational waves using laser interferometers with extreme sensitivity~\cite{LIGO2016}. At such precision, quantum fluctuations of light become the main source of noise. Two effects dominate: shot noise from photon counting statistics and radiation pressure from zero-point fluctuations of the electromagnetic field~\cite{Caves1980, Caves1981}. Recent experiments overcome these limits by injecting nonclassical light such as squeezed vacuum states~\cite{Abadie2011, LIGO2019, LIGO2024}. Squeezing redistributes quantum uncertainty between conjugate variables, so that enhanced photon number fluctuations $\langle\left(\Delta n_{\rm ph}\right)^2\rangle$ reduce phase fluctuations $\langle\left(\Delta\phi \right)^2\rangle$ and improve phase resolution. 

The fundamental limit on phase sensitivity is set by the quantum Fisher information~(QFI) 
${F}_{\rm Q}$, which saturates the Cram\'{e}r-Rao bound~\cite{Braunstein1994}:
\begin{equation}
1/\langle\left(\Delta\phi \right)^2\rangle \le M {F}_{\rm Q} ,    
\end{equation}
where $M$ is the number of measurements. The phase parameter is imprinted through the density matrix 
transformation $\rho(\phi) = e^{-i \phi O} \rho \, e^{i \phi O}$ generated by an operator $O$. The QFI quantifies how statistically distinct nearby states are in 
parameter space. For a pure state $\rho = |\psi_0 \rangle\langle\psi_0|$, it is proportional to the variance of the generator:
\begin{equation}
    {F}_{\rm Q} = 4\left(\langle O^2 \rangle_0  - \langle O 
\rangle_0^2 \right),
\end{equation}
where $\langle O \rangle_0 = \langle\psi_0| O|\psi_0 \rangle $. It follows that the QFI carries information about quantum correlations. Separable $N$-particle states exhibit classical scaling ${{F}_{\rm Q} \sim  N}$, while ${(k+1)}$-partite entanglement implies enhancement 
${{F}_{\rm Q} \sim k 
N}$, and maximally coherent superpositions reach the Heisenberg limit ${F}_{\rm Q} \sim N^2$~\cite{PhysRevA.85.022321, PhysRevA.85.022322, RevModPhys.90.035005}. 

When the generator $O$ is
the photon number operator $n_{\rm ph} = a^\dagger a$, the QFI density $f_{\rm Q} = F_{\rm Q}/\left(4\langle O 
\rangle_0\right)$ bears a close resemblance to the Fano factor from mesoscopic transport theory. The Fano factor compares the variance of a fluctuating quantity to its mean. When expressed in units of the elementary charge, it serves as the standard characteristic of shot 
noise, describing both classical and quantum fluctuations of electric currents~\cite{Levitov1996,BLANTER20001, PhysRevLett.96.246802, PhysRevB.76.085333}. Especially, it has been proposed that the quantum noise of the transport in a quantum point contact can be used to generate the quantum entanglement of electrons in the leads, and the noise offers the direct method to measure the entanglement entropy~\cite{PhysRevLett.102.100502}. A value $F=1$ manifests Poisson statistics while deviations (sub-Poissonian $F<1$ or super-Poissonian $F>1$) signal correlations between detection events~\cite{Scully_Zubairy_1997, walls2008quantum}. In this way, the QFI unifies the achievable precision of quantum measurements, fluctuations, and entanglement~\cite{Hauke2016}.

Continuous-variable systems are a natural platform for studying correlated photon states. For example, the two-mode squeezed vacuum state is a canonical realization of Einstein-Podolsky-Rosen correlations in the infinite-dimensional Hilbert space of a harmonic oscillator~\cite{Einstein1935, Braunstein2005}. Beyond Gaussian states, non-Gaussian states such as optical Schr\"{o}dinger cat and Gottesman-Kitaev-Preskill states have emerged as promising candidates for encoding qubits in bosonic modes~\cite{Gottesman2001, Takeda2019}. They have been engineered in microwave 
circuit quantum electrodynamics~\cite{Brock2025}, trapped ions~\cite{Fluhmann2019}, and only recently in the optical frequency regime~\cite{PhysRevX.13.031001, Konno2024}. As these previously exotic states become experimentally accessible, it is increasingly important to develop efficient methods for detecting their quantum properties, i.e., quantum coherence and many-body entanglement as quantified by the QFI.

Another of Einstein's fundamental contributions, the theory of the photoelectric effect, laid  
the interface between light and matter and originally postulated the quantum nature of light. 
In conventional photodetectors, however, a large photon flux is often used to improve the 
signal-to-noise ratio. In this high intensity limit, light effectively behaves as a coherent state, whereas quantum correlations are washed out. On the other hand, experiments with bright 
squeezed vacuum states suggest that large photon numbers and strong quantum 
correlations can still coexist~\cite{Gorlach2023}. 
This motivates a more general question: can the full counting statistics of a 
photocurrent reveal the QFI of the incident light? If 
so, which solid-state mechanism is most suited for this task?

For the manipulation, creation, and detection of quantum photons,
materials showing nonlinear optical responses are
indispensable~\cite{Boyd,Bloembergen,Sturman}.
Especially, the photovoltaic effect in noncentrosymmetric
materials has recently attracted intense interest. 
A representative example is the shift current, a 
second-order bulk photovoltaic response originating from the Berry 
phase of Bloch states~\cite{Baltz,Sipe,Young-Rappe,Morimoto2016}. 
When light excites an electron-hole pair, the pair experiences a 
real-space displacement that generates an electric polarization, whose relaxation produces a dc current~\cite{Belinicher1980, 1982JETP, PhysRevLett.133.206903}. 
Although its magnitude does not depend on the relaxation time, dissipation is 
necessary to sustain a steady current~\cite{Morimoto2016, PhysRevB.101.045201}. 
The shot noise of the shift current has also been studied 
assuming classical coherent light
\cite{PhysRevLett.121.267401}. It has been concluded that 
the shot noise as well as the external 
bias-dependent current come from the photocarriers. 
Due to its locality and geometrical origin, 
shift current is robust and has 
been proposed for efficient 
optoelectronic conversion. 

Shift current is similar in nature to the 
polarization current and does
not require photocarriers created by interband transitions. 
This feature enables shift current generation with photon energies 
less than the band gap, e.g., by excitations 
of excitons~\cite{PhysRevB.94.035117,Nakamura2024}, 
electromagnons~\cite{PhysRevB.100.235138,Ogino2024}, and 
phonons~\cite{Okamura22,p4s2-jpyk}. Exciton-polariton shift current has been 
proposed in the strong light-matter coupling regime~\cite{Morimoto2020}. 
In all these cases, no photocurrent shot noise occurs from photocarriers and hence the current noise is expected to reflect the
quantum nature of the photons. These unique properties of shift current make it promising 
for quantum photodetection. 

In this paper, we show that the shot noise of 
exciton-polariton photocurrents is highly sensitive to multiphoton 
entanglement and photon number coherence in the incoming light. By 
solving the coupled exciton-photon dissipative dynamics for an 
arbitrary initial photon density matrix, we find that the average 
shift current scales universally with the mean photon number and is independent of the specific form of the nonclassical light. While 
the average dc current is universal, the current fluctuations carry 
detailed quantum signatures. In particular, the Fano factor is determined by the 
photon number variance and directly encodes the QFI. These results are illustrated for optical cat states and squeezed 
vacuum states and highlight shift current statistics as a promising experimental method for 
characterizing quantum light beyond classical correlations.

\begin{figure*}
   \centering
   \includegraphics[width=0.99\linewidth]{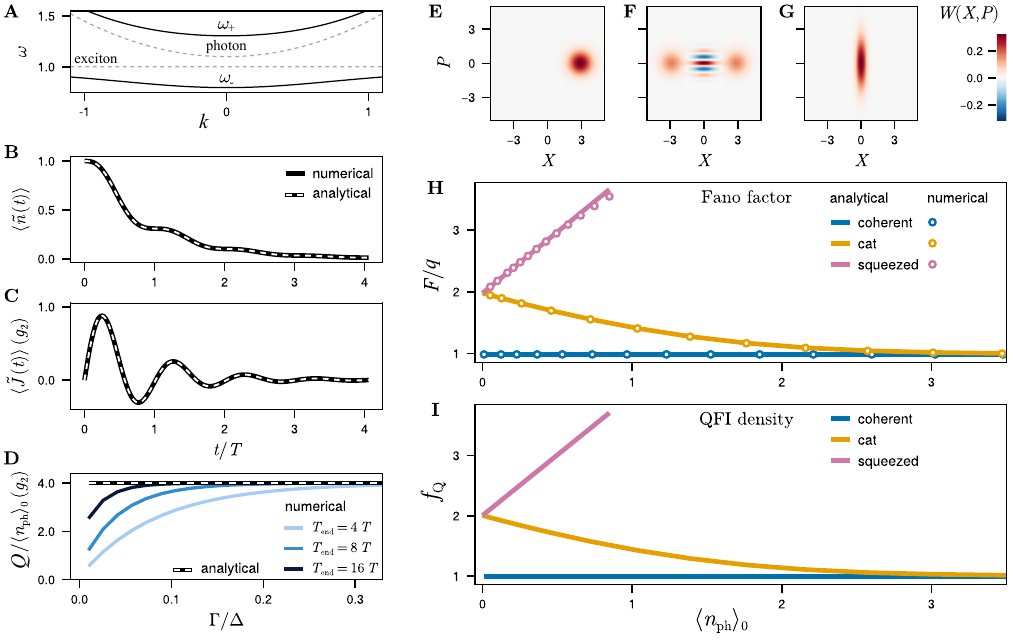}
   \caption{{Exciton polariton shift current and photon statistics.} 
   (A) Upper and lower polariton dispersion branches (solid lines). Dashed lines indicate the bare exciton and photon dispersions ($\omega_{\rm ex} = 1$ sets the energy unit, and $\omega_{\rm ph}=1.1$ at $k=0$). Parameters are $g_1 =0.25i$, a real $g_2$ that sets the current scale, and the damping constant $\Gamma=0.2$.
   (B),(C) Time evolution of the total particle number and the shift current for an initial single-photon state. Solid and dashed lines show numerically computed and analytical results, which coincide. 
   The time scale is set by $T={2\pi}/{\Delta}$, where $\Delta$~($\approx0.5$ for these parameters) is the polariton gap.
   (D) Time-integrated current as a function of $\Gamma$, calculated for an arbitrary initial photon state and normalized by the mean photon number $\langle n_{\rm ph} 
\rangle_0$. The dashed line shows the analytic single-photon response -- the shift charge $q=|g_2/g_1|$. Solid lines show numerical values that converge with increasing total simulation time $T_{\rm end}$.
   (E)-(G) Wigner functions in optical phase space $\alpha = X + i P$ for the coherent, Schr\"{o}dinger cat, and squeezed vacuum states. 
   (H)~Fano factor, normalized by $q$, as a function of the initial mean photon number for these states. Solid lines show analytic solutions; markers show numerical data. (I)~Corresponding quantum Fisher information density, 
   $f_{\rm Q} = F_{\rm Q}/\left(4\langle n_{\rm ph} 
\rangle_0\right)$. 
   } 
\label{fig:Fano}
\end{figure*}

\begin{table}[t]
\centering
\renewcommand{\arraystretch}{1.25}

\begin{tabular}{@{}p{0.5\linewidth} p{0.45\linewidth}@{}}
\toprule
Initial state $|\psi_0 \rangle$ & Definition \\
\midrule
Fock &
$\displaystyle
a^\dagger
\lvert 0 \rangle
$ \\[3pt]

Coherent &
$\displaystyle
\lvert \alpha \rangle=
\exp\!\left({\alpha a^\dagger - \alpha^* a}\right)
\lvert 0 \rangle
$ \\[3pt]

Optical Schr\"{o}dinger cat &
$\displaystyle
 \mathcal{N}_{\mathrm{cat}}
\big(\lvert \alpha \rangle + \lvert -\alpha \rangle\big)
$ \\[3pt]

Squeezed vacuum &
$\displaystyle
 \exp\!\left({r \left(aa - a^\dagger a^\dagger\right)/2 }\right)
\lvert 0 \rangle
$ \\[3pt]

Two-mode squeezed vacuum &
$\displaystyle
 \exp\!\left({r \left(a_1 a_2 - a_1^\dagger a_2^\dagger\right) }\right)
\lvert 0 \rangle$ \\
\bottomrule
\end{tabular}

\caption{
Single-photon state and representative continuous-variable quantum states~\cite{walls2008quantum}. $a^\dagger$ and $a$ are photon creation and annihilation operators.
$\mathcal{N}_{\mathrm{cat}}$ is a normalization constant, and
$r$ is the squeezing strength.
}\end{table}

\section*{Exciton polariton shift current} 
\subsection*{Model without dissipation}
\noindent
We consider the response of exciton polaritons to quantum light in a minimal cavity model~\cite{RevModPhys.82.1489}. The Hamiltonian describes a single photon mode coupled to a dispersionless exciton:
\begin{equation}
\label{eq:Ham0}
H_0 = \omega_{\rm ph} a^\dagger a + \omega_{\rm ex} b^\dagger b 
+ g_1^* a^\dagger b + g_1 b^\dagger a  \, .
\end{equation}
The bosonic operators $a^\dagger$ and $b^\dagger$ create a photon and an exciton with energies $\omega_{\rm ph}$ and $\omega_{\rm ex}$. The linear coupling $g_1$ represents the paramagnetic light-matter interaction. A typical 
polariton spectrum is shown in Fig.~\ref{fig:Fano}A. 

Following 
Ref.~\cite{Morimoto2020}, we include nonlinearity through the diamagnetic term
$H_{\rm int} = -A\left(g_2^* a^\dagger b + g_2 b^\dagger a\right)$, where $A$ is a static probe field. The associated current operator is obtained by differentiating with respect to $-A$: 
\begin{equation}
J= g_2^* 
a^\dagger b + g_2 b^\dagger a \, .
\end{equation}
Note that $g_2$ originally represents the interband diamagnetic interaction, which is related to the shift vector~\cite{Morimoto2016}.
Time-reversal symmetry fixes the relative phase of the couplings, $g_1/|g_1| = \pm ig_2/|g_2|$. Without loss of generality we use real $g_2$ and purely imaginary $g_1$, since any phase of $g_2$ can be removed by a global phase shift. 

In classical nonlinear optics, the shift current is a dc response quadratic in the applied field, $j_{\rm dc}=\sigma_{\rm shift} E(\omega)E(-\omega)$. In the present non-equilibrium dynamics, an initial photon excitation converts to the quantum shift current. Its time-integrated expectation value can be defined as a shift charge
\begin{equation}
\label{eq:shift_current}
    Q =  \int_{0}^\infty dt
    \langle \tilde{J}(t) \rangle,
\end{equation}
where Schr\"{o}dinger-picture operators $O$ are transformed to the interaction picture, 
$\tilde{O}(t) = e^{i H_0 t } O e^{-i H_0 t } $. The expectation value is given by $\langle \tilde{O}(t) \rangle = {\rm tr} \left(\tilde{O}(t) \tilde{\rho}(t)\right)$, where the density matrix in the interaction picture is $\tilde{\rho}(t) = e^{i H_0 t } \rho (t) e^{-i H_0 t }$.

Higher moments describe current fluctuations and are expected to be sensitive to quantum correlations. In analogy with mesoscopic transport, the Fano factor for shift current fluctuations can be defined as
\begin{equation}
    F = \frac{1}{Q}\int_{0}^\infty dt \int_{0}^\infty 
    d\tau  \left(C(t+\tau,t) + C(t,t+\tau)\right),
\end{equation}
where 
$    C(t_1,t_2) =
    \langle \tilde{J}(t_1) \tilde{J}(t_2) \rangle
    -\langle \tilde{J}(t_1) \rangle
    \langle \tilde{J}(t_2) \rangle$ is the connected current--current correlation function. 
The Fano factor is a promising observable for quantifying the QFI of the initial photon state, but it is ill-defined in the absence of losses. Without dissipation, the exciton-polariton dynamics is nothing but coherent Rabi oscillations, and hence the net current and other time-averaged observables vanish. 

The unperturbed Hamiltonian is readily diagonalized, owing to its resemblance to the Zeeman interaction. The creation operators evolve as 
\begin{equation}
\begin{split}
\tilde{a}^\dagger(t) &=  \phi_1(t) a^\dagger + \phi_2(t) b^\dagger   ,
\\
\tilde{b}^\dagger(t) &=  \chi_1(t)a^\dagger + \chi_2(t) b^\dagger,
\end{split}    
\end{equation}
where the coefficients are determined by the polariton energies and mixing angles (see Appendix~\ref{ap:A}). When the initial excited state contains only photons, the current expectation value is proportional to the mean photon number:
\begin{equation}
\label{eq:Fock1}
 \langle \tilde{J}(t) \rangle_0 = j_{0}(t)\, 
  \langle n_{\rm ph} \rangle_0,  \end{equation}
where $j_{0}(t) = g_2 \chi_1(t)\phi^*_1(t) + g_2 ^*\chi^*_1(t)\phi_1(t)$ is the current induced by a single photon in the Fock state. Here, the expectation value reduces to that in the Heisenberg picture, since the density matrix in the Schr\"{o}dinger picture exhibits only the coherent dynamics, i.e., $\tilde{\rho}(t) = \rho(0)$. Henceforth, we denote expectation values with respect to the density matrix in the initial state as $\langle \tilde{O}(t) \rangle_0= {\rm tr} \left(\tilde{O}(t) \rho(0)\right)$.

Likewise, the current correlator can be calculated,
\begin{equation} 
\label{eq:Fock2}
 \langle \tilde{J}(t_1) \tilde{J}(t_2) \rangle_0 = c_{\rm 0}(t_1,t_2)\, 
 \langle n_{\rm ph}^2 \rangle_0 + I_{\rm 0}(t_1,t_2)\, 
 \langle n_{\rm ph} \rangle_0, 
\end{equation}
where $c_{0}(t_1,t_2)=j_{0}(t_1)j_{0}(t_2)$, and 
$I_{0}(t_1,t_2)=j_{\rm int}(t_1)j^*_{\rm int}(t_2)$ with the interference matrix element 
$j_{\rm int}(t) = g_2^*\phi_1(t)\chi^*_2(t) + g_2\chi_1(t)\phi^*_2(t)$. These relations illustrate the coherent dynamics of the model, but are not generally applicable because their time integrals vanish. In the present case, the Fano factor relies on time-averaged observables, which are not well defined in the absence of losses.

\subsection*{Lindblad equation of dissipative dynamics}

\noindent 
Energy dissipation is crucial for obtaining finite time-integrated observables and supporting a net current generation.
Namely, within the time-reversible quantum mechanics, after a photon is absorbed and subsequently re-emitted, there is no mechanism to dissipate energy and generate a net current through relaxation of electric polarization.

For this reason, we introduce dissipation by coupling the exciton to an external bosonic bath. 
Integrating out the bath within the Markov approximation leads to the Lindblad equation, which describes the time evolution of the reduced density matrix~\cite{Lindblad1976, Gorini1976, gardiner2004quantum, Petruccione2007}:
\begin{equation}
\label{eq:Lindblad0}
\frac{\partial {\rho}(t)}{\partial t} = -i[H_0, \rho(t)] +\Gamma\left(
{b} {\rho}(t) {b}^\dagger - 
\frac{1}{2}\left\{ {b}^\dagger {b}, 
{\rho}(t) \right\} 
\right).
\end{equation}
The first term describes (von Neumann) coherent evolution, whereas the second accounts for loss into the bath. The damping constant $\Gamma$ governs the relaxation rate. The bath occupation is assumed to be negligible, which sets an effective zero-temperature limit with no stimulated absorption or emission. 

In the interaction picture, the evolution simplifies to the purely dissipative dynamics of the density matrix
\begin{equation}
\label{eq:Lindblad}
\frac{\partial \tilde{\rho}(t)}{\partial t} = \Gamma\left(
\tilde{b}(t) \tilde{\rho}(t) \tilde{b}^\dagger(t) - 
\frac{1}{2}\left\{ \tilde{b}^\dagger(t) \tilde{b}(t), 
\tilde{\rho}(t) \right\} 
\right).
\end{equation} 
The calculations are cumbersome, yet the physical picture behind this equation is simple: an initially 
decoupled photonic state hybridizes with the exciton and subsequently leaks coherently 
into the bath. Note that the excitons and photons are neutral and hence the leakage of the 
charge to the reservoir does not occur. The system is insulating, while it can support the 
net charge shift due to the geometric nature of the shift current. Our goal is to collect the total leakage current and characterize 
its quantum statistics.

\subsection*{Analytic solution}
\noindent
We study the relation between the Fano factor and the QFI for an arbitrary initial photon state. 
The state is represented in the basis of coherent states~\cite{PhysRev.131.2766, PhysRevLett.10.277}, 
\begin{equation}
\label{eq:Glauber}
\rho(0) = \int d^2\alpha P(\alpha) \ket{\alpha}\bra{\alpha}.
\end{equation} 
Here $P(\alpha)$ is the Sudarshan-Glauber quasiprobability distribution, and $\ket{\alpha}$ is a coherent state -- an eigenstate of the annihilation operator, $a\ket{\alpha}=\alpha\ket{\alpha}$. Coherent states are the closest quantum analog to a classical electromagnetic field, where $\alpha$ plays the role of the field's complex amplitude. This representation provides an intuitive phase-space picture that makes it easier to track quantum features of photon statistics.

Our main results are summarized in Fig.~\ref{fig:Fano}. We begin by deriving an analytic solution for the dissipative dynamics.
The Lindblad equation can be conveniently formulated in the superoperator language,
\begin{equation}
    \partial_t \tilde{\rho}(t) 
= \mathcal{\tilde{L}} (t) \tilde{\rho}(t),
\end{equation}
wherein the Liouvillian $\mathcal{\tilde{L}} (t)$ acts on the space of operators and density matrices 
rather than wavefunctions~\cite{gardiner2004quantum, Petruccione2007}. Its formal solution is  
$\tilde{\rho}(t) = 
\mathcal{S}(t,0) \, \rho(0)$, with the propagator given by the time-ordered exponential
\begin{equation} 
\mathcal{S}(t,0) = \mathcal{T} \exp\!\left({
\int_{0}^t dt' \mathcal{\tilde{L}} (t')}\right).
\end{equation}
If the system contained only one kind of boson, the evolution could be simplified by using the 
superoperators $\mathcal{K}_+ \rho = b \rho b^\dagger $, $\mathcal{K}_- \rho = b^\dagger \rho b $, 
$\mathcal{K}_z \rho = -\frac{1}{2}\left\{b^\dagger b + \frac{1}{2}, 
\rho\right\}$. These operators form an SU(1,1) algebra. Expressing the 
Liouvillian as a linear combination of them allows one to disentangle the exponential via Baker-Campbell-Hausdorff identities~\cite{PhysRevA.33.2444, Yurke1986, Ban1993}. 

In the present problem of two kinds of bosons, the Hilbert space is larger and there is a 
larger number of operators relevant to the solution (see Eq.~\eqref{eq:supops1} in Appendix~\ref{ap:B}). For 
example, $\mathcal{K}_+$ is generalized to 4 operators:
$\mathcal{K}_+^{1}\rho = a \rho a^\dagger$, 
$\mathcal{K}_+^{2}\rho = b \rho a^\dagger$, 
$\mathcal{K}_+^{3}\rho = a \rho b^\dagger$, and $\mathcal{K}_+^{4}\rho = b \rho b^\dagger$. 
A major simplification arises in the Sudarshan-Glauber representation. Namely, each photon trajectory $\ket{\alpha}\bra{\alpha}$ is an eigenstate of the operators ${\mathcal{K}}^a_+$. 
Acting on the initial density matrix, only one operator survives: 
$\mathcal{K}_+^{1}\ket{\alpha}\bra{\alpha} = \alpha \ket{\alpha}\bra{\alpha}\alpha^*$, 
while the others give exactly zero. This property becomes useful once the current operator 
is expressed in terms of the time-dependent  operators $\tilde{\mathcal{K}}^a_+(t)$. 

This can be achieved by evolving these superoperators in the 
dissipative setting. To do so, we introduce the dissipation picture by dressing interaction-picture superoperators with the dissipative propagator $\mathcal{S}(t,0)$:
\begin{equation}
\dtilde{\mathcal{O}}(t) = \mathcal{S}(0,t)\, \tilde{\mathcal{O}}(t)\, 
\mathcal{S}(t,0).    
\end{equation}
In the following, operators are first transformed into the interaction picture and subsequently evolved under the dissipative dynamics generated by $\mathcal{S}(t,0)$. The unitary transformation to the interaction picture is denoted by a tilde, while the non-unitary transformation to the dissipation picture is denoted by a double tilde.
The equation of motion then becomes
\begin{equation}
\label{eq:eqofmot}
\frac{d {\dtilde{{\mathcal{O}}}(t)}}{dt} = \bigl[\dtilde{\mathcal{O}}(t), \mathcal{\dtilde{L}} (t)\bigr] 
+ \frac{\partial \dtilde{\mathcal{O}}(t)}{\partial t} \, ,
\end{equation}
where ${\partial_t \dtilde{\mathcal{O}}(t)} = \mathcal{S}(0,t)\, \partial_t \tilde{\mathcal{O}}(t) \, 
\mathcal{S}(t,0)$.

Fortunately, one can access observables directly, without explicitly computing the Liouvillian or 
the density matrix in this representation. 
After inserting the identity $\mathcal{S}(t,0)\mathcal{S}(0,t)=1$, the mean current can be written as
\begin{equation}
\label{eq:meancurrent}
        \langle \tilde{J}(t) \rangle = 
    {\rm tr} \left( \tilde{J}(t) \tilde{\rho}(t)\right) 
    = {\rm tr} \left( \mathcal{S}(t,0) \dtilde{\mathcal{J}}(t)  \,  
    \rho(0)\right), 
\end{equation}
where $\dtilde{\mathcal{J}}(t)  = \mathcal{S}(0,t)\, \tilde{\mathcal{J}}(t)\, \mathcal{S}(t,0)$ is the current superoperator. 
Initially, $\tilde{{J}}(t)$ is rewritten as a superoperator $\tilde{\mathcal{J}}(t)$ by placing the creation operators to the right and the annihilation operators to the left of the density matrix $\tilde{\rho}(t)$. Such a procedure is well defined: the current operator is already normally ordered, and the rearrangement is allowed under the trace whereat it amounts to a cyclic permutation. Therefore, once found, the analytic solution for the mean current is exact.

We seek solutions for the current superoperator in the linear form
\begin{equation}    
\label{eq:ansatz}
\dtilde{\mathcal{J}}(t) = J_a\dtilde{\mathcal{K}}^a_+(t),
\end{equation}
described by the four-dimensional vector $J_a = (0,g^*_2,g_2,0)$ with summation over $a$ implied.
The key observation is that both the commutator and time derivative in Eq.~\eqref{eq:eqofmot} map the 
space of operators $\dtilde{\mathcal{K}}^a_+(t)$ onto itself. Their dynamics reduces to rotations 
in a four-dimensional operator space, analogous to the Landau-Lifshitz spin dynamics. 
The evolution can be expressed through matrices ${N}^a_b(t)$ such that 
$\dtilde{K}_+^a(t) = {N}^a_b(t) \tilde{K}_+^b(t)$. 

Once $\dtilde{\mathcal{J}}(t)$ is known, only ${K}_+^1$ contributes when it acts on the initial state. 
This operator can be isolated using $\tilde{K}_+^a(t) = M^a_b (t) {K}_+^b$ with 
coefficients $M^a_b (t)$ derived from $\chi_{1,2}(t)$ and $\phi_{1,2}(t)$ (see Appendix~\ref{ap:B}). 
The current expectation value then becomes
\begin{equation}
\label{eq:meancurrent1}
        \langle \tilde{J}(t) \rangle = j(t)\int d^2\alpha P(\alpha) |\alpha|^2,
\end{equation}
where $j(t) = J_a{N}^a_b(t)M^b_c(t) e^c$ with the unit vector $e^c=(1,0,0,0)$ selecting the action with ${K}_+^1$. 
Dissipation therefore adds 
damping into the coefficient $j_{0}(t)$ in Eq.~\eqref{eq:Fock1} and renders the time 
integral finite: 
\begin{equation}
\label{eq:JFock}
q = \int_{0}^\infty dt\,
    j(t). 
\end{equation}
The current integral is therefore universally proportional to the mean photon number,
\begin{equation}
Q = q\langle n_{\rm ph} \rangle_0,    
\end{equation}
independent of the nature of light in the initial state.

The same strategy works for the two-current correlator, but with an important caveat. The quantum regression theorem suggests~\cite{gardiner2004quantum, Petruccione2007}
\begin{equation} 
 \label{eq:varcurrent}
\begin{split}
 \langle \tilde{J}(t+\tau) \tilde{J}(t) \rangle &= \,
    {\rm tr} \left(\tilde{J}(t + \tau) \mathcal{S}(t+\tau,t)\tilde{J}(t) \tilde{\rho}(t)\right) 
    \\
    &=  {\rm tr} \left( \mathcal{S}(t+\tau,0) 
    \dtilde{\mathcal{J}}(t+\tau)\dtilde{\mathcal{J}}(t)  \,  
    \rho(0)\right) 
        \\& 
+{\rm interference} \ {\rm terms}.
    \end{split}
\end{equation} 
For the first contribution, the sequential action of the superoperators results in
\begin{equation} 
\label{eq:varcurrent1}
\begin{split}
    \langle \tilde{J}(t+\tau) \tilde{J}(t)\rangle &=
c(t+\tau, t)\int d^2\alpha P(\alpha) |\alpha|^4 \\&  + I(t+\tau, t),
\end{split}
\end{equation}
where $c(t+\tau, t)=j(t+\tau)j(t)$ and $I(t+\tau, t)$ represents the interference terms. As compared with 
Eq.~\eqref{eq:Fock2}, only the term proportional to $\langle a^\dagger a^\dagger a a \rangle_0$ survives. 

The remaining interference terms are a byproduct of rewriting both current operators as superoperators (see Appendix~\ref{ap:B}). The origin of these terms can be understood as follows. In Eq.\eqref{eq:varcurrent}, $\tilde{J}(t + \tau)$ can be immediately represented as the superoperator, as in the case of the mean current. By contrast, doing the same for $\tilde{J}(t)$ generates additional terms each time the creation operator from $\tilde{J}(t)$ commutes with one of the operators in $\dtilde{\mathcal{J}}(t + \tau) \mathcal{S}(t+\tau,t)$. As a consequence, while the analytic solution captures the contributions from the normal ordered part, it does not capture the interference terms originating from commutation relations. That said, the above expression is sufficient for obtaining the real part of its time average, and hence the Fano factor. This argument is further enforced by the sum rule that we will explain later.

\subsection*{Numerical solution}
\noindent 
The above analytical solution suggests that the integrated shift current depends only on the mean photon number. It is insensitive to other quantum properties of the photon states. By contrast, the Fano factor carries the QFI and thus captures the quantum features. 
In order to verify these results, we perform numerical simulations on the current response to quantum light. The dynamics are obtained numerically by integrating the Lindblad equation for the full photon-exciton density matrix in a truncated Fock space containing up to 15 particles in total. 


We focus on the response to the single-photon Fock state and three representative continuous-variable initial states: the coherent state, the optical Schr\"{o}dinger cat state, and the squeezed vacuum state. Their expressions are listed in Table~1. Their Wigner functions are well known in literature~\cite{walls2008quantum} and shown in Figs.~\ref{fig:Fano}E--G. The coherent state has a Gaussian Wigner function that resembles a localized wavepacket in optical phase space $\alpha = X + iP$. It may be viewed as a displaced vacuum and saturates the minimum uncertainty relation $\Delta X\Delta P =1/2 $ ($\hbar=1$). Because of this property, it corresponds to a classical optical field and thus serves as a natural classical reference. By contrast, the squeezed vacuum and optical cat states exhibit nonclassical features. The squeezed state is also Gaussian but anisotropic: a reduction of $\Delta X$ is accompanied by an increase in $\Delta P$, such that fluctuations in the $X$ quadrature are below those in the vacuum. This behavior has no classical analog. The optical cat state is intrinsically non-Gaussian, especially for small number of photons, as featured by interference fringes that arise from quantum coherence between two closely placed coherent states.

We first examine the response for an initial single-photon state. The time evolution of the total particle number and the 
shift current is shown in Figs.~\ref{fig:Fano}B and C. The numerical values agree with the analytic prediction 
in Eq.~\eqref{eq:meancurrent1}. Figure~\ref{fig:Fano}D shows the integrated current for several values of the damping constant $\Gamma$ at different total simulation times. In all cases, the numerical curves converge to $q$, the universal coefficient introduced in Eq.~\eqref{eq:JFock} as the shift charge generated by a single photon. 

It follows that the shift charge $Q$ depends on the model parameters but not on $\Gamma$. In spite of that, finite dissipation is still necessary because the point $\Gamma=0$ is singular. This behavior is similar to the classical-field case of the two-band Floquet model, where the shift current scales as $\Gamma/\sqrt{A^2 + \Gamma^2}$, with a phenomenological parameter $A$ proportional to the applied field~\cite{Morimoto2016}. In our model, the limit $A\to 0$ is built in by construction, and a finite $\Gamma$ works as a physically meaningful regularization parameter for time integrals. 

Next, we examine the three representative continuous-variable states. After rescaling by the initial particle number, the current time dependence for all three representative states becomes identical to that for the single-photon state shown in Fig.~\ref{fig:Fano}C. It follows that the integrated current, i.e., the total shift charge $Q$, exhibits a universal linear 
dependence on the mean photon number (see Fig.~\ref{fig:Fano}D), in full agreement with the analytic prediction 
in Eq.~\eqref{eq:meancurrent1}.

Unlike the mean current, the fluctuations are sensitive to quantum correlations. Figures~\ref{fig:Fano}H and I compare the Fano factor of the shift current response with the QFI for the three representative states. The numerical Fano factors reproduce the analytical curves obtained from Eq.~\eqref{eq:varcurrent1}. Note that the analytical expression for the time dependence $C(t+\tau,t)$ misses a small $\tau$-oscillatory offset observed in the numerical results (see Appendix~\ref{ap:B} and Fig.~\ref{fig:Cjj}). Nevertheless, it accurately captures the time integrals of the real part, and hence the Fano factors. 
Since the current originates from nonlinear interactions, the Fano factor is naturally expressed in units of $g_2$. Up to a model-dependent prefactor, the numerical Fano factors scale linearly with the corresponding QFI density.

\section*{Sum rules and QFI}
\noindent 
It is remarkable that the Fano factors fully mirror the QFI density. This is unlikely to be accidental, but instead points toward more general conservation laws or sum rules, that remain valid even in the presence of exciton losses. 
Such sum rules indeed exist for all time-ordered correlators. 
They follow from a simple continuity relation for particle operators in the interaction picture:
\begin{equation}
    \tilde{J}(t)=-q\partial_t \left(\tilde{a}^\dagger(t) \tilde{a}(t)  \right) = q\partial_t \left(\tilde{b}^\dagger(t) \tilde{b}(t)  \right).
\end{equation}
Here $q = |g_2/g_1|$, the same quantity that appears in the shift current in Eq.~\eqref{eq:JFock}. 

This equivalence, as well as the identity $Q = q\langle n_{\rm ph} \rangle_0$, becomes clear by using the continuity relation to replace the current operator with the time derivative $\partial_t\tilde{n}_{\rm ph}(t)$ inside the expectation value $\langle \tilde{J}(t) \rangle$, and then integrating by parts (see Appendix~\ref{ap:C}).
After time integration, only the boundary term survives.
One can infer that the response is strongest 
when the nonlinear diamagnetic coupling $g_2$ dominates over the linear paramagnetic coupling $g_1$.

In a similar way, replacing both the current operators in the two-point correlator $\langle \tilde{J}(t + \tau) \tilde{J}(t) \rangle$ yields a sum rule for its real part, as detailed in Appendix~\ref{ap:C}. This leads to the exact relation for the Fano factor:
\begin{equation}
\label{eq:sum_rule_Fano}
F =  \frac{q}{\langle n_{\rm ph} \rangle_0}\biggl[\langle n_{\rm ph}^2 \rangle_0 - \langle n_{\rm ph} \rangle_0^2 \biggr] \,.    
\end{equation}
The sum rule is in agreement with calculations displayed in Fig.~\ref{fig:Fano}H, and comparing with the definition of the QFI density, one concludes the relation
\begin{equation}
\label{eq:sum_rule_Fano1}
F = q f_{\rm Q}  \,.    
\end{equation}

Together with analytical predictions and numerical solutions, this sum rule confirms that shift current shot noise can be used to track quantum correlations in many-body photon states. Coherent states, with QFI density 
\begin{equation}
{f}_{{\rm Q}}[{\rm coh}] = 1 ,    
\end{equation}
provide the classical reference level (see Fig.~\ref{fig:Fano}I). Nonclassical states behave differently. For optical cat states with mean photon numbers of order unity, the Fano factor captures quantum interference patterns. Their origin lies in the QFI of the cat state, with density 
\begin{equation}    
\label{eq:QFIcat}
{f}_{{\rm Q}}[{\rm cat}] 
= 1 + {4 s_{\mathrm{cat}} |\alpha|^2  }/\left({1-s_{\mathrm{cat}}^2}\right),  
\end{equation}
where $s_{\mathrm{cat}} = \bra{\alpha} \cdot \ket{-\alpha}= e^{-2|\alpha|^2}    
$ is the overlap between two nearby coherent states which renders the cat state nonclassical (see Appendix~\ref{ap:D}).
Similarly, in the case of squeezed vacuum states the numerical Fano factor follows the analytical expression for the QFI density
\begin{equation}
{f}_{{\rm Q}}[{\rm SV}] = 2 (\langle n_{\rm ph}\rangle_0 + 1),
\end{equation}
where the mean particle number is $\langle n_{\rm ph}\rangle_0 = \sinh^2{r}$. 
In the limit of small photon numbers, cat states approach the QFI of a bipartite entangled state, 
whereas squeezed vacuum states indicate a crossover from classical scaling 
${f}_{{\rm Q}}[{\rm SV}]\sim 1$ to Heisenberg scaling ${f}_{{\rm Q}}[{\rm SV}] \sim \langle n_{\rm ph}\rangle_0$ 
at larger photon fluxes.

\section*{Discussion}
\noindent
We have proposed a new quantum photodetection scheme using shift current of exciton polaritons. Upon a photon excitation, exciton carries no charge and only generates nonlinear photocurrent via shift mechanism. The integrated current depends only on the number of photons in the initial state, $Q = q\langle n_{\rm ph} \rangle_0 $. Another viewpoint is that the accumulated shift charge is quantized in units of $q=|g_2/g_1|$, which is the shift charge induced by a single photon. 

\section*{Conclusion}
\noindent
Nonlinear optical responses have been developed rapidly 
by the invent of lasers, which produce coherent
and intense light. In these cases, the light can often be 
regarded as classical and described by a coherent state.
On the other hand, advances in photodetection have enabled the detection of small numbers of photons
and hence the nonlinear optical processes driven 
by quantum photon states~\cite{PhysRevX.13.031001, Konno2024, RevModPhys.88.045008, Tsuji2025}.
Our results demonstrate a measurable link between shift current statistics and quantum correlations of light. In particular, the nonlinear photocurrent of exciton polaritons converts photon number fluctuations into measurable shot noise, so its Fano factor directly reflects the QFI encoded in the incident quantum light. Materials that host quantum-geometric shift current, when integrated with optical cavities, may therefore provide a practical instrument to assist continuous-variable quantum metrological applications~\cite{yokomizo2026} and quantum computation with bosonic codes~\cite{Gottesman2001, Takeda2019}.

\section*{Acknowledgments}
\noindent
We thank Ying-Ming Xie and Koki Shinada for fruitful discussions. E.B. is sincerely grateful to Ying-Ming Xie, Chunli Huang and Aaron Merlin M\"uller for their critical feedback on the manuscript. N.N. and E.B. were supported by JSPS KAKENHI Grant Numbers 24H00197, 24H02231 and 24K00583. T.M. was supported by JSPS KAKENHI Grant 23K25816, 23K17665, and 24H02231. N.N. was supported by the RIKEN TRIP initiative.

\appendix
\section{Exciton polariton model}
\label{ap:A}
\noindent

\begin{figure*}
   \centering
   \includegraphics[width=0.99\linewidth]{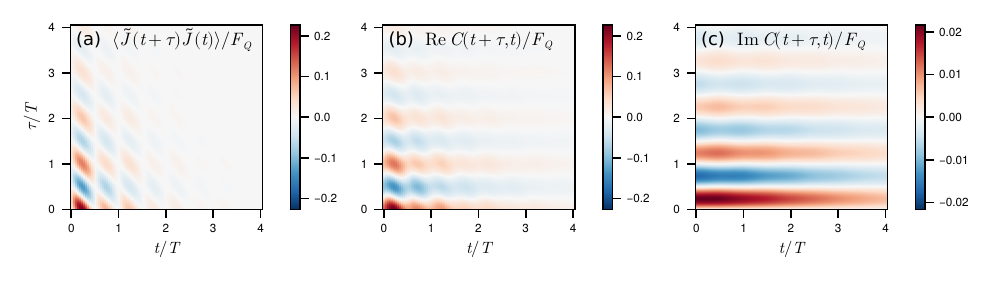}
   \caption{Time dependencies of current-current corelation functions. (a) Analytical result normalized with $F_{\rm Q}$. (b), (c) real and imaginary part of the numerical counterpart. Here $\Gamma=0.1$ and the other model parameters are the same as those in Fig.~\ref{fig:Fano}. The initial state is the optical cat state with $\alpha =0.5$.
   } 
\label{fig:Cjj}
\end{figure*}

\noindent

\noindent
The unperturbed Hamiltonian can be written as
\begin{equation}
\label{eq:Ham0ap}
H_0 =\left(a^\dagger, b^\dagger \right) \begin{pmatrix}
\omega_{\rm ph} & g_1^*\\
g_1 & \omega_{\rm ex}
\end{pmatrix}  \begin{pmatrix}
a\\
b
\end{pmatrix} =h_0 I + \bm{h} \cdot \bm{\sigma},
\end{equation}
with $h_0 = \frac{\omega_{\rm ph} + \omega_{\rm ex}}{2}$, $h_z = 
\frac{\omega_{\rm ph} - \omega_{\rm ex}}{2}$, $h_x = {\rm Re}(g)$, 
and $h_y = {\rm Im}(g)$. The boson doublet is now described by Pauli matrices $\bm \sigma$ and identity matrix $I$. $H_0$ becomes diagonal in a rotated frame where the pseudospin aligns with the effective field direction 
$
\bm n_h =\bm h /|\bm h| = \left(\sin \theta\cos \phi,\sin \theta \sin \phi ,  \cos \theta\right).
$
The resulting eigenmodes are the upper and lower polaritons
\begin{equation}
\begin{pmatrix}
c_+^\dagger\\
c_-^\dagger
\end{pmatrix} =\begin{pmatrix}
\cos\frac{\theta}{2} & e^{i\phi} \sin\frac{\theta}{2}\\
-\sin\frac{\theta}{2} & e^{i\phi} \cos\frac{\theta}{2}
\end{pmatrix} 
\begin{pmatrix}
a^\dagger\\
b^\dagger
\end{pmatrix},
\end{equation}
with energies
\begin{equation}
E_\pm = h_0 \pm |\bm h| = \frac{\omega_{\rm ph} + \omega_{\rm ex}}{2}  
\pm \sqrt{ \left(\frac{\omega_{\rm ph} - \omega_{\rm ex}}{2}\right)^2 + |g|^2}.
\end{equation}
The polariton operators evolve as 
$\tilde{c}_{\pm}^\dagger(t) = e^{i E_\pm t } c_{\pm}^\dagger$. Hence the operators $a^\dagger$ and $b^\dagger$ in the interaction picture are described by the coefficients $\phi_\sigma(t)$ and $\chi_\sigma(t)$:
\begin{equation}
\begin{split}
\phi_1(t) &= \cos^2\!\left(\frac{\theta}{2}\right) 
e^{i E_+ t } + \sin^2\!\left(\frac{\theta}{2}\right) e^{i E_- t }\, ,\\
\phi_2(t) &= e^{i\phi} \sin \theta \frac{e^{i E_+ t } 
- e^{i E_- t }}{2} \, ,
\end{split}
\end{equation} 
and
\begin{equation}
\begin{split}
\chi_1(t) &= e^{-i\phi} \sin \theta \frac{e^{i E_+ t } 
- e^{i E_- t }}{2} \, ,\\
\chi_2(t) &= \sin^2\!\left(\frac{\theta}{2}\right) 
e^{i E_+ t } + \cos^2\!\left(\frac{\theta}{2}\right) e^{i E_- t }\, .
\end{split}
\end{equation} 
They form orthonormalized pseudospinors, provided by $\phi^\dagger(t)\phi(t) = 1$, $\chi^\dagger(t)\chi(t) = 1$, and $\phi^\dagger(t)\chi(t) = 0$.
In the pseudospin notation $c_1= a$ and $c_2= b$, and  
$\tilde{c}_1^\dagger(t) =  \phi_\sigma(t) c^\dagger_\sigma$ and $\tilde{c}_2^\dagger(t) =  \chi_\sigma(t) c^\dagger_\sigma$ with summation over $\sigma=1,2$ implied.

\section{Analytic solution of Lindlbad equations}
\label{ap:B}
\noindent 
The single-boson superoperators form an SU(1,1) algebra with commutation relations $[\mathcal{K}_z, \mathcal{K}_\pm] = \pm \mathcal{K}_\pm$ and $[\mathcal{K}_+, \mathcal{K}_-] = -2 \mathcal{K}_z$. By analogy, we construct a set of polariton operators:
\begin{equation}
\label{eq:supops}
\begin{split}
\mathcal{K}_+^a \rho &= \sigma^a_{\sigma\sigma'}c_{\sigma'} \rho c_{\sigma}^\dagger \\
\mathcal{K}_z^a \rho & = -\frac{1}{2}\sigma^a_{\sigma\sigma'}\left\{c_{\sigma}^\dagger c_{\sigma'}  + \frac{1}{2}I_{\sigma\sigma'}, \rho\right\}, \\
\end{split}
\end{equation}
where $\sigma^a=(I, \bm \sigma)$ is the Pauli vector in four dimensions. In the present basis $a=0,+,-,z$ and $\sigma_{\pm} = \sigma_x \pm i\sigma_y$. The explicit structure is as follows:
\begin{equation}
\label{eq:supops1}
\begin{split}
\mathcal{K}_+^0 \rho &= c_1 \rho c_1^\dagger + c_2 \rho c_2^\dagger \\
\mathcal{K}_+^+ \rho &= 2 c_2 \rho c_1^\dagger\\
\mathcal{K}_+^- \rho &= 2 c_1 \rho c_2^\dagger\\
\mathcal{K}_+^z \rho &= c_1 \rho c_1^\dagger - c_2 \rho c_2^\dagger\\
\end{split}
\quad
\begin{split}
\mathcal{K}_z^0 \rho & = -\frac{1}{2}\left\{c_1^\dagger c_1 
+ c_2^\dagger c_2 + 1, \rho\right\}  \\
\mathcal{K}_z^+ \rho & = -\frac{1}{2}\left\{2c_1^\dagger c_2 , \rho\right\} \\
\mathcal{K}_z^- \rho & = -\frac{1}{2}\left\{2c_2^\dagger c_1 , \rho\right\}\\
\mathcal{K}_z^z \rho & = -\frac{1}{2}\left\{c_1^\dagger c_1 
- c_2^\dagger c_2 , \rho\right\}. \\
\end{split}
\end{equation}
Their commutation relations are: $\bigl[\mathcal{K}_+^a, \mathcal{K}_+^b \bigr] = 0$ and 
$\bigl[\mathcal{K}_z^a, \mathcal{K}_+^b \bigr] = f^{ab}_c \mathcal{K}_+^c$, 
with nonzero structure constants $f^{+-}_0=f^{-+}_0=2$ together with $f^{00}_0, f^{0z}_z,f^{z0}_z,f^{zz}_0, f^{0+}_+, f^{0-}_-,f^{+0}_+,f^{-0}_-$ equal to unity. These relations hold in both the interaction and dissipation pictures only at equal times, e.g., for $\tilde{\mathcal{K}}_+^a(t)$ and $\dtilde{\mathcal{K}}_+^a(t)$. 

The Liouvillian can then be written as
\begin{equation}
    \tilde{\mathcal{{L}}} (t) =  d_a \left(\tilde{\mathcal{K}}_{+}^a(t) + 
    \tilde{\mathcal{K}}_{z}^a(t)\right),
\end{equation}
where the only nonzero coefficients are $d_0=\Gamma/2$ and $d_z=-\Gamma/2$. Here the Euclidian scalar product is $\sum_{a=0,+,-,z}X_a Y^a = X^0 Y^0 + X^z Y^z + \left(X^+ Y^- + X^- Y^+ \right)/2$. Substituting operators into Eq.~\eqref{eq:eqofmot} leads to the equation of motion
\begin{equation}
\frac{d \dtilde{\mathcal{K}}_{+}^a(t)}{dt} = - D^a_c\dtilde{\mathcal{K}}_{+}^c(t)
+ \frac{\partial \dtilde{\mathcal{K}}_{+}^a(t)}{\partial t} \, ,
\end{equation}
where $D^a_c = f^{ab}_c d_b$. Its nonzero matrix elements are $D^0_0 = D^+_+ = D^-_- = D^z_z = \Gamma/2$ and $D^0_z  = D^z_0 = -\Gamma/2$. The coherent part of the evolution is given by
\begin{equation}
{\partial_t {\tilde{\mathcal{K}}_+^a(t)}} = i H^{a}_b \tilde{\mathcal{K}}_+^b(t),
\end{equation}
where the only nonzero elements are $H^z_- =g_1$, $H^z_+ =-g^*_1$, $H^+_z =-2 g_1$, $H^-_z =2 g^*_1$, ${H^+_+ = \omega_{\rm ph} - \omega_{\rm ex}}$, ${H^-_- = -\omega_{\rm ph} + \omega_{\rm ex}}$. 
We consider a solution of the form
\begin{equation}
\tilde{\tilde{\mathcal{K}}}_+^a(t) = {N}^a_b(t) \tilde{\mathcal{K}}_+^b(t)
\end{equation}
with initial conditions ${N}^a_b(0) = \delta^a_b$.
Collecting all coefficients of independent operators gives
\begin{equation}
{\partial_t {{N}^a_d(t)}} 
= -D^a_c {N}^c_d(t) 
+ i\left( H^{a}_b {N}^b_d(t) - {N}^a_b(t) H^{b}_d  
\right).
\end{equation}
In matrix form $\partial_t {N}(t) = -D {N}(t) +i (H {N}(t) - {N}(t)H)
$ such a differential equation can be readily solved 
$
{N}(t) = e^{-Dt+iHt} {N}(0) e^{-i H t}
$.
Extracting the identity part of $D$, the solution is
\begin{equation}
N^a_c(t) = e^{-\Gamma t/2}\left(e^{\bar{D}t+iHt} \right)^a_b \left(e^{-i H t}\right)^b_c\, ,
\end{equation}
where the only nonzero elements of $\bar{D}$ are $\bar{D}^0_z  = \bar{D}^z_0 = \Gamma/2$. The dissipation leads to a boost-like transformation in the $a = 0,z$ sector, while the coherent Hamiltonian dynamics generates pure rotations within the three-dimensional subspace $a = x,y,z$. 


This analytic solution is exact only for mean observables. The difficulty is in representing the current operator by the superoperator $\dtilde{\mathcal{J}}(t)$ in Eq.~\eqref{eq:ansatz}. For mean values this mapping becomes one-to-one correspondence because the trace is cyclic. For example, ${\rm tr}\left(\tilde{a}^\dagger(t) \tilde{b}(t) \, \tilde{\rho}(t) \right) = {\rm tr}\left(\tilde{b}(t) \, \tilde{\rho}(t) \tilde{a}^\dagger(t)\right) = {\rm tr}\left(\tilde{\mathcal{K}}_+^{+}(t) \, \tilde{\rho}(t) \right)/2 $. Using this identity the current expectation value is expressed as the action of $\mathcal{K}_+^a$. 

This argument also applies for higher-order correlators, but with missing interference terms. Creation and annihilation operators at different times do not commute, and they also do not commute with the intermediate evolution operator, such as $\mathcal{S}(t+\tau,t)$ in Eq.~\eqref{eq:varcurrent}.
The interference terms occur due to these commutations. The commutation relations for operators at different times are
\begin{equation}
\begin{split}
[\tilde{a}(t'), \tilde{a}^\dagger(t)] &= \phi^\dagger(t')\phi(t), 
\quad
[\tilde{b}(t'), \tilde{b}^\dagger(t)] = \chi^\dagger(t')\chi(t), 
\end{split}
\end{equation}
and
\begin{equation}
\begin{split}
[\tilde{a}(t'), \tilde{b}^\dagger(t)] &= \phi^\dagger(t')\chi(t), 
\quad
[\tilde{b}(t'), \tilde{a}^\dagger(t)] = \chi^\dagger(t')\phi(t).
\end{split}
\end{equation}
Due to time-translational symmetry of the unperturbed polariton model, these commutators are functions of the time difference $\tau=t'-t$.
Thus the two-point correlator $C(t+\tau,t)$ shows oscillations in $\tau$, as shown in Fig.~\ref{fig:Cjj}.
Nervertheless, the present analytic method correctly reproduces the real part of the total time integrals of $C(t+\tau,t)$, as shown by the sum rule in Eq.~\eqref{eq:sum_rule_Fano} and further confirmed by the numerical simulation in Fig.~\ref{fig:Fano}.
We note that these commutators give a finite contribution to the imaginary part of the time integral.

As an example, we calculate the current expectation value in Eq.~\eqref{eq:meancurrent1}. In the Pauli basis, the transverse vector is $J_a = (0,g_2^*,g_2,0)/2$, and the light-cone vector is $e^c =(1,0,0,1)$. To separate the action of $\mathcal{K}_+^1$ onto the initial density matrix, it remains to obtain $M^a_b(t)$ explicitly. In the interaction picture this action is
\begin{equation}
\begin{split}
    \tilde{\mathcal{K}}_+^a(t) \rho(0) &= \sigma^a_{\sigma_1\sigma_2}\tilde{c}_{\sigma_2}(t) \rho(0) \tilde{c}_{\sigma_1}^\dagger(t) \\&
=
\sigma^a_{\sigma_1\sigma_2}U^*_{\sigma_2\sigma_2'}(t) U_{\sigma_1\sigma_1'}(t){c}_{\sigma_2'} \rho(0) {c}_{\sigma_1'}^\dagger \, ,
\end{split}
\end{equation}
Here $U_{\sigma \sigma'}$ is a bi-spinor such that $U_{1\sigma'}(t) = \phi_{\sigma'}(t)$ and $U_{2\sigma'}(t) = \chi_{\sigma'}(t)$. 
The relevant matrix elements are: $M^0_a(t) e^a=|\phi_1(t)|^2+|\chi_1(t)|^2 $, $M^z_a(t) e^a=|\phi_1(t)|^2 - |\chi_1(t)|^2 $, $M^+_a(t) e^a=2\phi_1(t) \chi_1^*(t) $, and $M^-_a(t) e^a=2\phi_1^*(t) \chi_1(t) $. This finalizes the calculation of the mean current. The total particle number can be obtained by formally replacing $J_a\to (1,0,0,0)$. Higher-order correlators are computed by analogy. 

\section{Sum rules}
\label{ap:C}
\noindent
In the interaction picture operators satisfy the following equation of motion
\begin{equation}
    \partial_t \tilde{O}(t) = ie^{iH_0t} [H_0, O] e^{-iH_0t}.
\end{equation}
For the photon number operator $n_{\rm ph}=a^\dagger a$, using the commutator $[H_0,a^\dagger a]= g_1b^\dagger a-g_1^*a^\dagger b$, the continuity relation writes as
\begin{equation}
    \partial_t \tilde{n}_{\rm ph}(t) = i\left(g_1 \tilde{b}^\dagger(t)\tilde{a}(t) - g_1^* \tilde{a}^\dagger(t)\tilde{b}(t) \right).
\end{equation}
When the couplings choice is $g_1^*=-g_1$ and $g_2^*=g_2$, this expression simplifies to 
\begin{equation}
\tilde{J}(t)=-q \,\partial_t\tilde{n}_{\rm ph}(t)  ,   
\end{equation}
where we denote the coefficient $q \equiv|g_2/g_1|$. 

In the case of the mean current in Eq.~\eqref{eq:meancurrent}, the sum rule is as follows.
Substituting the continuity relation into the current expectation value and integrating it by parts, we obtain the shift charge
\begin{equation}
\label{ap:eq:shift}
    Q = -q\int_0^\infty dt \,{\rm tr}\biggl[  \partial_t  \left( \tilde{n}_{\rm ph}(t)\tilde{\rho}(t)\right) 
    -   \tilde{n}_{\rm ph}(t)\partial_t\tilde{\rho}(t)
    \biggr].
\end{equation}
The second term vanishes. This comes from the Lindblad equation of motion; see Eq.~\eqref{eq:Lindblad}. Applying it, 
\begin{equation}
\label{ap:eq:trace_perm}
\begin{split}
        \frac{2}{\Gamma}{\rm tr}\biggl( \tilde{n}_{\rm ph}(t)\partial_t\tilde{\rho}(t)
    \biggr)=&
    2\,{\rm tr}\biggl( \tilde{a}^\dagger(t)\tilde{a}(t)\tilde{b}(t) \tilde{\rho}(t) \tilde{b}^\dagger(t)
    \biggr)\\    
    -&{\rm tr}\biggl( \tilde{a}^\dagger(t)\tilde{a}(t)\tilde{b}^\dagger(t) \tilde{b}(t)\tilde{\rho}(t)
    \biggr)\\
    -&
    {\rm tr}\biggl( \tilde{a}^\dagger(t)\tilde{a}
    \tilde{\rho}(t)(t)\tilde{b}^\dagger(t) \tilde{b}(t)
    \biggr)
    .
    \end{split}
\end{equation}
These creation and annihilation operators are at equal times and obey canonical commutation relations and thus can be simply permuted under the trace. 
Integrating the first term in Eq.~\eqref{ap:eq:shift} gives the shift charge
\begin{equation}
    Q  = q \, {\rm tr} \left( \tilde{n}_{\rm ph}(0)\tilde{\rho}(0)\right) 
    .
\end{equation}
This exactly reproduces Eq.~\eqref{eq:JFock}, confirming that $Q/\langle n_{\rm ph} \rangle_0 = |g_2/g_1| =q$.

In the case of the two-current correlators, the sum rule is as follows. Integrating by parts in $\tau$ gives
\begin{equation} 
\label{ap:eq:two_cor}
\begin{split}
& 
\int_{0}^\infty 
d\tau\,
{\rm tr} \left(\tilde{J}(t + \tau) \mathcal{S}(t+\tau,t)\tilde{X}(t)\right) =
\\&
-q
\int_{0}^\infty 
d\tau \, {\rm tr}\biggl[ \partial_\tau\bigl(\tilde{n}_{\rm ph}(t + \tau) \mathcal{S}(t+\tau,t)\tilde{X}(t)\bigr) 
\\
&
\qquad \qquad \qquad \qquad
-\tilde{n}_{\rm ph}(t + \tau) \partial_\tau\mathcal{S}(t+\tau,t)\tilde{X}(t)
\biggr]
\end{split}
\end{equation} 
where for brevity we introduced an operator ${\tilde{X}(t)} = {\tilde{J}(t) \tilde{\rho}(t)}$. 
After substituting the equation of motion for the evolution superoperator, ${\partial_\tau\mathcal{S}(t+\tau,t)} = {\mathcal{\tilde{L}}(t+\tau)\mathcal{S}(t+\tau,t)}$, the second term becomes
\begin{equation}
\label{ap:eq:trace_perm1}
\begin{split}
        \frac{2}{\Gamma}{\rm tr}\biggl( \tilde{n}_{\rm ph}(t')\mathcal{\tilde{L}}(t') (...)
    \biggr)=&
    2\,{\rm tr}\biggl( \tilde{a}^\dagger(t')\tilde{a}(t')\tilde{b}(t') (...) \tilde{b}^\dagger(t')
    \biggr)\\    
    -&{\rm tr}\biggl( \tilde{a}^\dagger(t')\tilde{a}(t')\tilde{b}^\dagger(t') \tilde{b}(t')(...)
    \biggr)\\
    -&
    {\rm tr}\biggl( \tilde{a}^\dagger(t)\tilde{a}(t)
    (...)\tilde{b}^\dagger(t) \tilde{b}(t)
    \biggr),
    \end{split}
\end{equation}
where we denoted $t' =t + \tau$ and replaced with $(...)$ a whatever operator comes out as the result of acting $\mathcal{S}(t+\tau,t)$ onto $\tilde{X}(t)$.
Thus, the second term is zero as before; compare with Eq.~\eqref{ap:eq:trace_perm}. 

The first term in Eq.~\eqref{ap:eq:two_cor}, after integrating in $\tau$, is equal to $q\,{\rm tr}\biggl[ \tilde{n}_{\rm ph}(t ) \tilde{X}(t) \biggr]$. It can be summed up with the complex conjugated term $q\,{\rm tr}\biggl[ \tilde{X}^\dagger(t) \tilde{n}_{\rm ph}(t ) \biggr]$, where $\tilde{X}^\dagger(t) = \tilde{\rho}(t) \tilde{J}(t)$. Their $t$-integral is then
\begin{equation}
    q\int_{0}^\infty 
dt \, {\rm tr}\biggl[ \left(
\tilde{n}_{\rm ph}(t ) \tilde{J}(t) + \tilde{J}(t) \tilde{n}_{\rm ph}(t )
\right)\tilde{\rho}(t) \biggr]
\end{equation}
After substituting the continuity relation for the remaining current operator, one can notice that the sum of the resulting operators is the total derivative $\partial_t\left(\tilde{n}^2_{\rm ph}(t )  \right)$. So that the time integral again reduces to a boundary term, as in the case of the mean current. 

Finally, the sum rule reads as
\begin{equation}
    F = \frac{q^2}{Q} \biggl[ {\rm tr} \left( \tilde{n}^2_{\rm ph}(0)\tilde{\rho}(0)\right)
    - \bigl[{\rm tr} \left( \tilde{n}_{\rm ph}(0)\tilde{\rho}(0)\right)\bigl]^2
    \biggr] ,
\end{equation}
where the second term comes from subtraction of $ Q^2$ that completes the disconnected correlator.
This results in Eq.~\eqref{eq:sum_rule_Fano} and in the main text.

\section{QFI of Schr\"{o}dinger cat state}
\label{ap:D}
\noindent
The variance of the photon number is
\begin{equation}
    {\rm Var} \left( n_{\rm ph} \right) =\langle n_{\rm ph}^2 \rangle_0 - \langle n_{\rm ph} \rangle_0^2= I_2 + I_4 - I_2^2 \, .
\end{equation}
Here $I_2=\langle n_{\rm ph} \rangle_0 = \langle a^\dagger a \rangle_0$ is the mean photon number, and $I_4= \langle a^\dagger a^\dagger a a \rangle_0$. They come from rewriting the operators in normal order. Using the Sudarshan-Glauber representation in Eq.~\eqref{eq:Glauber},
\begin{equation}
    I_2 =  \int d^2\alpha P(\alpha) |\alpha|^2 , \quad
    \quad
    I_4 =  \int d^2\alpha P(\alpha) |\alpha|^4.
\end{equation}
For such a photon density matrix, we need to calculate $n$-th order matrix elements for arbitrary coherent states described by complex numbers $\alpha_{\rm in}$ and $\alpha_{\rm fin}$:
\begin{equation}
\label{eq:ap:D_matel}
    \langle \alpha_{\rm fin} | \left(a^\dagger\right)^n \left(a\right)^n |\alpha_{\rm in}\rangle =
    \left(\alpha_{\rm fin}^*\right)^n\left(\alpha_{\rm in}\right)^n\langle \alpha_{\rm fin} | \cdot |\alpha_{\rm in}\rangle,
\end{equation}
where the overlap between the coherent states (see Ref.~\cite{gardiner2004quantum} for details) is given by
\begin{equation}
\langle\alpha_{\rm fin} | \cdot |\alpha_{\rm in}\rangle= \exp\!\left(\alpha_{\rm fin}^*\alpha_{\rm in} -|\alpha_{\rm fin}|^2/2 - |\alpha_{\rm in}|^2/2 \right)    .
\end{equation}

For a coherent state, the density matrix is $\rho[{\rm coh}] = |\alpha\rangle \langle\alpha|$.
The number of particles in the coherent state follows Poisson statistics, where the variance of the photon number operator is equal to its mean. This originates from $I_2 = |\alpha|^2$ and $I_4=I_2^2 =|\alpha|^4$, resulting from Eq.~\eqref{eq:ap:D_matel}. Hence the QFI density, which is given by ${\rm Var} \left( n_{\rm ph} \right)/I_2$, is equal to 1.

In the case of the Schr\"{o}dinger cat state, there occurs an additional contribution due to its nonclassical features. As a result, $I_4\ne I^2_2$. The density matrix is 
\begin{equation}
\begin{split}
    \rho[{\rm cat}] = \frac{1}{\mathcal{N}}&\biggl(
|\alpha\rangle \langle\alpha|+
|-\alpha\rangle \langle-\alpha| \\ &
\qquad
+
|\alpha\rangle \langle-\alpha|+
|-\alpha\rangle \langle\alpha|
\biggr),
\end{split}
\end{equation}
where $\mathcal{N} = 2\left(1 + s_{\mathrm{cat}}\right)$ with $s_{\mathrm{cat}} = \bra{\alpha} \cdot \ket{-\alpha}= e^{-2|\alpha|^2}    
$.
Then the matrix elements are 
$I_2={|\alpha|^2(1-s_{\mathrm{cat}})/(1+s_{\mathrm{cat}})}$ and $I_4 ={|\alpha|^4}$. Therefore the nonclassical correction becomes ${I_4-I_2^2}={4|\alpha|^4s_{\mathrm{cat}}/(1+s_{\mathrm{cat}})^2}$.   
Consequently, the variance device from unity by the following factor
\begin{equation}
    \left(I_4-I_2^2\right)/I_2 = {4 s_{\mathrm{cat}} |\alpha|^2  }/\left({1-s_{\mathrm{cat}}^2}\right).
\end{equation}
It leads directly to the QFI density given in Eq.~\eqref{eq:QFIcat} in the main text. Finally the mean and variance for squeezed vacuum states is known in literature~\cite{Braunstein2005, gardiner2004quantum}.

\bibliographystyle{apsrev4-2}
\bibliography{ref.bib}

\end{document}